\documentclass[onecolumn]{svjour3}          
\usepackage{graphicx}
\usepackage{adjustbox}
\usepackage{amsmath}
\usepackage{epstopdf}
\usepackage[graphicx]{realboxes}
\usepackage{rotating}
\usepackage{float}
\usepackage{latexsym}
\usepackage{amssymb}
\usepackage{array}
\usepackage{longtable}
\usepackage{dcolumn}
\usepackage{bm}

\begin{document}

\title{ Nonsingular Black Holes in Higher dimensions }

\author{Bikash Chandra Paul}

\institute{ B. C. Paul \\
\email{bcpaul@nbu.ac.in}\at
Department of Physics, University of North Bengal, Siliguri, Dist. : Darjeeling 734 013, West Bengal, India \\
            and \\
            IUCAA Centre for Astronomy Research and Development, North Bengal  }

\date{Received: date / Accepted: date}

\maketitle

\vspace{10pt}

\begin{abstract}
We present a class of new nonsingular  black holes in  higher dimensional theories of gravity. Assuming a specific form of the stress energy tensor exact analytic solutions of the field equation are generated in general theory of relativity (GR) and Rastall theory. The non-singular black hole solutions are obtained with a finite pressure at the centre in $D=4$ dimensions. For $D>4$ the transverse pressure is found finite at the centre for a set of  model parameters. In the later case the transverse pressure is more than that in the usual four dimensions. The exact analytic solution of the  field equations in higher dimensions for large $r$ coincides with the Schwarzschild black hole solution in the usual four and in higher dimensions which is singularity free. The different features of the generalized non-singular black hole in GR and modified GR  are explored. A new vacuum nonsingular black hole is found in Rastall gravity. We also study the motion of massive and massless particles around the black holes.
\end{abstract}

\maketitle	
%\vspace{0.2cm}

\section{ Introduction}
\label{intro}

The idea that spacetime dimensions should be extended from four to higher dimensions came
from the seminal work of Kaluza and Klein \cite{1,2} who first tried to unify gravity with electromagnetism.
The Kaluza-Klein approach has been revived and considerably generalized after realizing that many
interesting theories of particle interactions need spacetime dimensions more than four for their 
formulation. During the last few decades considerable research  activities  in progress to understand the quantum properties of gravity.  The investigation seems to lead some people to believe that a consistent theory of quantum gravity cannot be obtained within the framework of point-field theories. For example, superstring theory is considered to be the promising candidate which may unify gravity with the other fundamental forces in nature which requires ten dimensions for consistent formulation. The advent of string theory has opened up new and interesting
possibilities in this context. The  discovery that a supergravity theory coupled to Yang-Mills fields with a gauge group SO(32) or   $ E_8$ $\times E_8$ is anomaly-free in ten dimensions had inspired considerable activities in this area.  Although the expected breakthrough has not yet come, worldwide hectic activities have served to focus on a number of issues which need further investigations in higher dimensions.
 The present ideas in dimensional reduction suggest that our cosmos may be a 3-brane evolving in a D-dimensional spacetime. 
Cadeau and Woolgar \cite{3}  addressed this issue in the context of black holes which led to homogeneous but non-
FRW-braneworld cosmologies.  
Classical General Relativity (GR) in more than the usual four dimensions is thus a subject of increasing
attention in recent years.  A successful development of counting   of  the  five dimensional black hole entropy \cite{3a} and the AdS/CFT correspondence relates the properties of a  $D$-dimensional black hole with those
of a quantum field theory in $(D-1)$ dimensions \cite{3b}. There has been a growing interest to investigate the physics of higher-dimensional black holes \cite{real} which is markedly
different, and much richer in structure compared to  four dimensions.

In connection with localized sources, higher dimensional generalization of the spherically symmetric Schwarzschild,
Reisner-Nordstr$\ddot{o}$m black holes, Kerr black holes can be  found in the literature \cite{4,5,6,7,8,9}. The  generalization of the rotating Kerr black hole \cite{5,6,10,11} and  black holes in compactified spacetime \cite{4,9} are also found in the literature. The linearized stability of the black holes  \cite{12},  no hair theorems \cite{13}, black hole thermodynamics and Hawking radiation  have also  been investigated. 
 Mandelbrot \cite{14} investigted the problem on the variability of
dimensions and  describe how a ball of thin thread is seen as an observer changes scale.
An object which looks like  a point object from a very large distance becomes a three-dimensional
ball visible at a  closer distance. Therefore at various scales the ball
appears to change shape as an observer moves down. While the embedding dimensions for the ball has not changed, the
effective dimensions of the contents  however also remains same. It is possible that there are compact \cite{15} or
non-compact \cite{16} dimensions present at a certain point. In this case  the (3 + 1) metric is simply not true, although one obtains a valid description with general relativity.  We also probe the black hole solution in Rastall theory \cite{R0} which is prescribed by a modification of GR to accommodate the present accelerating universe \cite{R1a,R2,R3}.

 Regular ($i.e.$ non-singular) black holes have  been initiated  by  Bardeen \cite{bar} and thereafter a number of black hole solutions  in four dimensions have been obtained \cite{18,19,20,21,21a,21b,21c,21d,21e}. In this case one
can find metrics which are spherically symmetric, static, asymptotically flat, with regular centres, and for which
the resulting Einstein tensor is physically reasonable, satisfying the weak energy condition and having components
which are bounded and fall off appropriately at large distance.
Dymnikova \cite{18} obtained nonsingular Schwarzsclid black hole solution in vacuum and thereafter extended to obtain nonsingular cosmological black hole \cite{19} solutions to include the de Sitter solution in the usual four dimensions. Formation and evaporation of non-singular black hole is also discussed \cite{sa} from an initial vacuum region accommodating Bardeen-like static region supported by finite density and pressures, subsequently its pressure vanishes rapidly at large radius which however behaves as a   cosmological constant at a small radius.
In 2019, Event Horizon Telescope group captured the first ever image of a supermassive black hole at the centre of the M87 galaxy which triggers the various possibilities for the state of the compact object and opened up new horizon in theoretical physics.
The motivation of the paper is to obtain a nonsingular black hole solution in higher dimensions and  investigate the different features of such black holes in GR and beyond GR. 
For this we consider Higher dimensional Einstein gravity (GR) and Rastall gravity for a comparative study.

  The paper is organised as follows: In sec. 2, the Einstein field equation  in a static higher dimensional metric is obtained.  In sec. 3,   non-singular black holes are obtained in Rastall theory with extension of spacetime dimensions. In sec.  4, we present analytical set up of the non-singular black hole solution to investigate the shadow of the black hole. The effective potential and  the shadow behaviour  of the black holes are analyzed in sec. 5. Finally we summarize in  sec 6.

\section{Einstein Field Equation in Higher Dimensions}

We consider a  higher dimensional gravitational action which is given by
\begin{equation}
\label{e1}
I= - \frac{1}{16 \pi G_D} \; \int \sqrt{-g} \;d^Dx  \;R +I_m
\end{equation}
where $R$ is the Ricci scalar, $G_D$ is the $D$-dimensional gravitational constant and $I_m$ represents the matter action.
The Einstein field equation is given by 
\begin{equation}
\label{e2}
R_{AB} -\frac{1}{2} g_{AB}R =  \kappa^2 T_{AB}
\end{equation}
where $A, B = 0,1, ... D-1$ and  $T^{A}_{B }= ( -\rho, P_r,  P_{\perp}, ...)$ the energy-momentum tensor, $\rho$ the energy density, $P_r$ the radial  pressure, $P_{\perp}$ transverse pressure, $\kappa^2=\frac{8 \pi G_D}{c^2}$.  We consider  $D$ dimensional spacetime metric given by
\begin{equation}
\label{3}
ds^{2}= -  e^{\nu} dt^{2} + e^{\lambda}  \;dr^{2}  + r^2 d\Omega^2_{D-2} 
\end{equation}
where $\nu$ and $\lambda$ are functions of radial coordinate $r$ and $\Omega_{D-2} $ is for unit sphere in $S^{D-2}$ dimensions.  
The components of the Einstein equations and the metric given by eq. (\ref{3}) are given by
\begin{equation}
\label{4}
T^{t}_{t} =\frac{(D-2)}{2} \left[  e^{-\lambda} \left( \frac{D-3}{r^2} - \frac{\lambda'}{r} \right) - \frac{D-3}{r^2}  \right],
\end{equation}
\begin{equation}
\label{5}
  T^{r}_{r} =\frac{(D-2)}{2} \left[ e^{-\lambda} \left( \frac{D-3}{r^2} + \frac{\nu'}{r} \right) - \frac{D-3}{r^2}  \right],
\end{equation}
\[
T^{\theta_1}_{\theta_1} =  -\frac{(D-3)(D-4)}{2r^2} +\frac{e^{-\lambda}}{2} \times
 \]   
 \begin{equation}
\label{6}
\left[ \nu'' +\frac{\nu'^2}{2} - \frac{\lambda' \nu'}{2} + \frac{(D-3)(\nu'-\lambda')}{r} + \frac{(D-3)(D-4)}{r^2}  \right] ,
\end{equation}
\begin{equation}
\label{7}
T^{\theta_2}_{\theta_2} =  ... ... = T^{\theta_{D-2}}_{\theta_{D-2}} =T^{\theta_1}_{\theta_1} 
\end{equation}
for simplicity we have taken $\kappa^2= \frac{8\pi G_{D}}{c^{2}}=1$. The radial null vector $l^A$ can be selected to have the components $l^t= e^{\lambda/2} $, $l^r= \pm e^{\nu/2} $ and  $l^i= 0 $. The two radial null-null components of the Ricci tensor are equal, and given by $R_{AB} l^A l^B= e^{\lambda}R_{tt} +e^{\nu} R_{rr}= (D-2) \frac{(e^{\lambda+\nu})'}{2 e^{\lambda+\nu}}$, which vanishes if and only if $(\lambda+\nu)$ is a constant. A rescaling of the  time coordinate can be set to make the sum of the terms equal to zero for black hole solution and  we write 
 %We consider here the following linear relation \cite{17} to look for black hole solutions 
% \begin{equation}
%\label{8}
%T^{t}_{t}  =T^{r}_{r} \Rightarrow \lambda' + \nu' =0.
%\end{equation}
%On integrating we get
%\begin{equation}
%\label{9}
% \lambda + \nu=g(t).
%\end{equation}
\begin{equation}
\label{10}
 \lambda + \nu=0
\end{equation}
Now substituting
\begin{equation}
\label{11}
 \lambda =- \ln f(r)
\end{equation}
we obtain the following components of energy momentum tensors in terms of  $f(r)$, which are given by 
\begin{equation}
\label{12}
T^{t}_{t}=T^{r}_{r} = \frac{D-2}{2}\left[ f(r) \left(\frac{D-3}{r^2} + \frac{f'}{rf(r)} \right) - \frac{D-3}{r^2} \right]
\end{equation}
\[
T^{\theta_1}_{\theta_1} = \frac{f(r)}{2}\left[ \frac{f''}{f} + \frac{2(D-3) f'}{rf(r)} + \frac{(D-3)(D-4) }{r^2}  \right] 
\]
\begin{equation}
\label{13}
\hspace{1.5 cm}  - \frac{(D-3)(D-4)}{2r^2} 
\end{equation}

\begin{equation}
\label{14}
T^{\theta_2}_{\theta_2} =  ... ... = T^{\theta_{D-2}}_{\theta_{D-2}} =T^{\theta_1}_{\theta_1} 
\end{equation}
where the prime denotes the derivative with respect to $r$.
The source term satisfying 
\begin{equation}
\label{15}
T^t_t =T^r_r, \; \; \;  and \;\;\; T^{\theta_2}_{\theta_2} =  ... ... = T^{\theta_{D-2}}_{\theta_{D-2}} =T^{\theta_1}_{\theta_1} 
\end{equation}
and the equation of state, $T_{B}^{A};_{A}=0$,. 

 Assume the density profile in higher dimensions $T^t_t=-\rho$ as
\begin{equation}
\label{17}
\rho=-T^t_t=\rho_0 \; e^{- \frac{r^{D-1}}{r_{*}^{D-1}}}
\end{equation} 
where $r_{*}$ is a dimensional constant connected with a constant density $\rho_0$. The density $\rho_0$ also permits a $D$ dimensional  de Sitter solution with its size given by 
\begin{equation}
r_0^2= \frac{(D-1)(D-2)}{2\Lambda},
\end{equation}
where $\Lambda=\rho_0$. Using the density profile given eq. (\ref{17}) in eq. (\ref{12})  we integrate and   obtain the metric potential which yields
\begin{equation}
\label{18}
f(r) = 1- \frac{r_g^{D-3}}{r^{D-3}} + \frac{ 2 \rho_0 r_{*}^{D-1}}{(D-1)(D-2) } \frac{1}{r^{D-3}} e^{- \frac{r^{D-1}}{r_{*}^{D-1}}}
\end{equation}
where $r_g^{D-3}=  \left( \frac{2 \rho_0 }{(D-1)(D-2)} \right) r_{*}^{D-1}$.
The higher dimensional metric is now can be written as 
\begin{equation}
\label{19}
ds^{2}= -  \left( 1- \frac{R_s(r)}{r^{D-3}}  \right) dt^{2} + \frac{ dr^{2}}{\left( 1- \frac{R_s(r)}{r^{D-3}}  \right) }  
 + r^2 d\Omega^{2}_{D-2}
\end{equation}
where we denote
\begin{equation}
\label{20}
R_s(r) = r_g^{D-3} \left[ 1 - exp \left(  -\frac{ r^{D-1}}{r_{*}^{D-1}} \right)\right]
\end{equation}
and 
\begin{equation}
\label{21}
r_{*}^{D-1} = r_0^{2} \; r_g^{D-3},
\end{equation}
where $r_0^2= \frac{(D-1)(D-2)}{2 \rho_0}$.
This is an exact spherically symmetric solution of the Einstein field equations in $D$-dimensions. For  $D=4$ the solution given by eq. (\ref{19})  reduces to the solution obtained by Dymnikova \cite{18}. The other components of energy momentum tensor can be obtained using the Einstein's field equations which are given by
\begin{equation}
\label{22}
T^{\theta_2}_{\theta_2}=  ... = T^{\theta_{D-2}}_{\theta_{D-2}}=   \left[ \frac{D-1}{D-2} \left(\frac{r}{r_{*}}\right)^{D-1} 
- 1\right] \rho_0 e^{- \frac{r^{D-1}}{r_{*}^{D-1}}}.
\end{equation} 
It is evident that in the usual 4 dimensions Dymnikova \cite{18} black hole solutions recovered with anisotropic fluid distribution when $r=r_{*}$, which is true also in higher dimensions.  The nonsingular black hole  (NSBH) solutions are  permitted with anisotropic fluid distributions in higher dimensions. The  energy density and radial pressure follow the vacuum configuration but the tangential pressures  do not. The tangential pressure is non-zero which remains  positive definite for $ r>r_{*}$. At the center the  tangential pressure is negative indicating existence of exotic matter ($P_{\perp} <0$) at the center of the black hole.  The nonsingular black hole solution obtained by Dymnikova can not be described in   lower dimension $D=2+1$, however, we can  extend the concept of NSBH  in more than the usual four dimensions. The generalization of the black hole solution in higher dimensions accommodates  a new class of NSBH solutions    where the tangential  pressure increases to a large extent inside the non-singular black hole with a different feature but away from the centre of the black hole it decreases exponentially. 

The mass of a massive object in higher dimension is given by
\begin{equation}
\label{v1}
m(r)=A_{D-2} \int^{r}_{0} r'^{D-2} \rho(r') dr'
\end{equation}
where $A_{D-2}=\frac{2\pi^{\frac{D-1}{2}}}{\Gamma(\frac{D-1}{2})}$ which at $r\rightarrow \infty $ is connected to the whole mass $\mathcal{M}$ connected with $r_g^{D-3}$  by the Schwarzschild relation.  The modulus difference between $R_s(r_g) $ and $r_g^{D-3}$ is  $r_g^{D-3} e^{- \frac{r^{D-1}}{r_{*}^{D-1}}}$. 
The  measure of the difference between the higher dimensional Schwarzschild  mass $m(r) \sim r_g^{D-3} $ in a singular black hole and $R_s$ of a non-singular black hole  is given by
\begin{equation}
\label{23}
 \frac{   \mathcal{M}   -m(r)}{\mathcal{M}} = exp \left( - \frac{r^{D-1}}{r_{*}^{D-1}} \right).
\end{equation} 
Here $m(r)$  becomes  $\mathcal{M}$ at infinite distance.
It is found that the mass difference decreases as the dimension in which black hole embedded increases. The metric has two event horizons located at 
\[
 r_{ +}  = r_g  \left[ 1- O\left(exp \left( -\frac{r_g^2}{r_0^2} \right) \right) \right], 
 \]
\begin{equation}
\label{24}
  r_{ -}  = r_0  \left[ 1- O\left(exp \left( -\frac{r_0}{r_g} \right) \right) \right].
\end{equation} 
Here $r_{+}$ is the external event horizon. The metric evaluated at $g_{tt} (r_{+} )=0$ describes an object with the similar properties properties of a black hole by a distant observer, it does not send light signals outside and could not interact with its surroundings by the gravitational field. In four dimensions it is found that both $r_{+} $ and $r_{-}$ are removable singularities of the metric. The singularities can be eliminated by an appropriate transformation.

\section{Higher Dimensional Rastall gravity}

In this section we explore  NSBH solution in the Rastall theory of gravity for $D\geq 4$ dimensions. The Rastall theory \cite{R0} is based on the modification of the Einstein field equation  for a spacetime with Ricci scalar filled by an energy momentum source as follows:
\begin{equation}
\label{r1}
T^{AB};A= \lambda R^B
\end{equation}
where $\lambda$ is the Rastall parameter which is a measure for deviation from the standard GR conservation law. Consequently the Rastall field equation can be written as
\begin{equation}
\label{r2}
G_{AB} + \kappa^2 \lambda  g_{AB} R= \kappa^2 T_{AB}
\end{equation}
 where $\kappa^2$ is the Rastall gravitational constant. The above field equation reduces to that of GR in the limit $\lambda \rightarrow 0$ and $\kappa^2=8 \pi G$. 
However, for a vanishing trace of the energy-momentum tensor, for example the electrovacuum solution can be obtained when $\lambda = \frac{1}{4}$ or $R=0$. It is important to note that the former possibility is
not physically acceptable as the trace of the energy momentum tensor   vanishes $T=0$ for any scalar field. 
Consequently the matter configuration where the energy-momentum tensor has
null trace, the relativistic solution obtained in Rastall theory is same as that one obtains in 
the general theory of 
relativity (GR). This feature of Rastall theory which is a modified GR  led us to look for  black holes solutions in a background of matter/energy with   non-vanishing
trace.  It may be pointed out here that  the Rastall gravity is widely used to accommodate 
acceptable explanation for the current acceleration of the universe which has no solution in  GR and for this it is interesting to explore NSBH in Rastall theory.

 We consider the metric for black hole solution in higher dimensions $D\geq 4$:
 \begin{equation}
\label{3}
ds^{2}= -  f(r) dt^{2} + \frac{dr^{2}}{f(r)}  + r^2 d\Omega^2_{D-2} .
\end{equation}
Using the metric, we obtain the non-vanishing components of the  Rastall tensor  $H_{AB} =G_{AB} +  \lambda  g_{AB} R$ and $\kappa^2=1$,
\begin{equation}
\label{r4}
H^{t}_{t} =  \frac{D-2}{2r^2} \left( r f'-(D-3) + (D-3)f   \right)      +  \lambda  R,
\end{equation}
\begin{equation}
\label{r5}
H^{r}_{r} =  \frac{D-2}{2r^2} \left( r f'-(D-3) + (D-3)f   \right)      +  \lambda  R,
\end{equation} 
 \begin{equation}
\label{r6}
H^{\theta_i}_{\theta_i} =  \frac{r^2 f'' +(D-3) (2 r f' + (D-4) (f -1) }{2r^2}     +  \lambda   R
\end{equation}
where $i=1, 2, ... , (D-2)$, and the Ricci scalar in $D$ dimensions is given by
\begin{equation}
\label{r7}
R =  - \frac{1}{r^2} \left( r^2 f'' +2 (D-2) r f' + (D-2)(D-3) (f -1)  \right)      
\end{equation}
in the above we denote $()'$ to represent derivative with respect to the radial coordinate $r$. We solve the field equation to obtain higher dimensional non-singular black holes in Rastall theory and for this $H^{t}_{t}=T^{t}_{t}$ and
 $H^{r}_{r}=T^{r}_{r}$ yield
\[
P_r= \frac{D-2}{2r^2} \left( r f'-(D-3) + (D-3)f   \right) 
 \]
 \begin{equation}
\label{r8}
 - \frac{\lambda }{r^2} \left( r^2 f'' +2 (D-2) r f' + (D-2)(D-3) (f -1)  \right),
\end{equation} 
 and also we consider
 $H^{\theta_1}_{\theta_1}=T^{\theta_1}_{\theta_1}$ , ... and  $H^{\theta_{D-2}}_{\theta_{D-2}} =T^{\theta_{D-2}}_{\theta_{D-2}}$ which yield
 \[
P_{\perp}=  \frac{1}{2r^2} \left( r^2 f''+ 2 (D-3) r f'  + (D-3) (D-4) (f-1) \right) 
\]
 \begin{equation}
\label{r8}
 - \frac{\lambda }{r^2} \left( r^2 f'' +2 (D-2) r f' + (D-2)(D-3) (f -1)  \right) .
\end{equation} 
In this case we explore the non-singular Black hole obtained in higher dimensional Rastall gravity, the general solution of the metric is
\begin{equation}
\label{r9}
f(r) = 1- \frac{r_g^{D-3}}{r^{D-3}} + \frac{ 2 \rho_0 r_{*}^{D-1}}{(D-1)(D-2) } \frac{1}{r^{D-3}} e^{- \frac{r^{D-1}}{r_{*}^{D-1}}}
\end{equation}
The energy density and radial pressure are
\begin{equation}
\label{r10}
\rho=  \left(  \frac{  D-2-2 \lambda D + 2 (D-1) \lambda \frac{r^{D-1}}{r_{*}^{D-1}}}{D-2} \right) \rho_0 e^{- \frac{r^{D-1}}{r_{*}^{D-1}}},
\end{equation}

\begin{equation}
\label{r11}
P_r= - \left(  \frac{  D-2-2 \lambda D + 2 (D-1) \lambda \frac{r^{D-1}}{r_{*}^{D-1}}}{D-2} \right) \rho_0 e^{- \frac{r^{D-1}}{r_{*}^{D-1}}},
\end{equation}
the tangential pressure is given by
\begin{equation}
\label{r12}
P_{\perp} = \left( (1-2 \lambda) \frac{D-1}{D-2} \frac{r^{D-1}}{r_{*}^{D-1}} -  \frac{D-2- 2 \lambda D }{D-2 } \right) \rho_0 e^{- \frac{r^{D-1}}{r_*^{D-1}}}.
\end{equation}
The energy density and the transverse pressure in Rastall gravity framework obtained    in eqs. (\ref{r10}) and  (\ref{r12}) reduces to the eqs. (\ref{17}) and (\ref{22}) in GR for $\lambda \rightarrow 0$. The modification introduced in GR by Rastall admits nonsingular Dymnikova \cite{18} black hole (NSBH) with normal matter while the radial pressure corresponds to vacuum equation of state.
 At the center  of the NSBH the energy density is $\rho= (D-2-2 \lambda D) \rho_0$, which increases as the. number of spacetime dimension increases for a given range of Rastall parameter $\lambda < \frac{D-2}{2D}$. It is evident that for a given dimension, NSBH admits greater mass for lower values of  $\lambda$ and the lower limiting value for  $\lambda < \frac{D-2}{2D}$ and   $|\lambda| >  \frac{D-2}{2D}$ (for negative $\lambda$). The corresponding tangential pressure at the center   $P_{\perp} = (2D\lambda+2-D)\rho_0$ is  finite but negative. 
 In $D=4$ dimensions, at the center of the black hole, $\rho (r=0)= 2(1-4 \lambda) \rho_0$ and tangential pressure $P_{\perp}= -(1-4\lambda) \rho_0$ which indicates existence of exotic matter at the centre in GR (as $\lambda=0$) as well as in Rastall theory for $\lambda > - \frac{1}{4}$.   Thus NSBH can be realized with both central radial pressure and tangential pressure  negative and equal but an anisotropy in pressure develops away from the center in Rastall gravity,  normal matter exists when $r> \left(  \frac{ D-2 -2 \lambda D}{(D-1) (1-2\lambda} \right)^{1/(D-1)} r_{*}$.  
  The tangential pressure indicates black hole  surrounded by exotic matter in Rastall gravity \cite{R1} for the range $\frac{1}{4} < \lambda < \frac{1}{2}$..
For $r\rightarrow \infty $, the energy density and pressure vanishes asymptotically.\\

 When $\lambda= \frac{D-2}{2D}$, we get the following :
 \begin{equation}
\label{r10a}
\rho=  \rho_0    \left( \frac{  D-1}{D} \frac{r^{D-1}}{r_{*}^{D-1}}\right) e^{- \frac{r^{D-1}}{r_{*}^{D-1}}},
\end{equation}

\begin{equation}
\label{r11a}
P_r= - \rho=  \rho_0    \left( \frac{  D-1}{D} \frac{r^{D-1}}{r_{*}^{D-1}}\right) e^{- \frac{r^{D-1}}{r_{*}^{D-1}}},\end{equation}
the tangential pressure is given by
\begin{equation}
\label{r12a}
P_{\perp} =2 \rho_0   \left( \frac{D-1}{D(D-2)} \frac{r^{D-1}}{r_{*}^{D-1}}  \right) 
 e^{- \frac{r^{D-1}}{r_*^{D-1}}}.
\end{equation} 
  one obtains NSBH with $\rho >0$, $\rho+P_r=0$ and $P_{\perp} > 0 $. 
   For $D=4$ dimensions, $\lambda=\frac{1}{4}$ and the NSBH  can be realized in Rastall gravity with normal matter which however is not permitted in GR.  Also we note that at the centre of the NSBH the tangential pressure vanishes, admitting a perfect vacuum NSBH in the usual four dimensions. The result obtained in this case is also applicable in higher dimensions. This is a new result.

  %when  $r < \left(\frac{D-2+2\lambda D}{(1-2 \lambda)(D-2)} \right)^{\frac{1}{D-1}}  r_{*}$. 

\section{ Analytical set up}
The modified Schwarzschild metric for a non-singular black hole is given by
\begin{equation}
\label{25}
ds^{2}= - f(r) \; dt^{2} + f(r)^{-1} \;  dr^{2}
 + r^2 d\Omega^{2}_{D-2}
\end{equation}
where $f(r) = 1-  \left( \frac{r_g}{r} \right)^{D-3} +   \left( \frac{r_g}{r} \right)^{D-3} \;     exp \left[ - \left(  \frac{r}{r_{*}}\right)^{D-1} \right]$,
making use of the assumption   $\kappa^2=8 \pi $ made earlier, we write $r_g=\left( \frac{16 \pi M}{(D-2) A_{D-2}}\right)^{\frac{1}{D-3}}$  and the area of $D$ dimensional sphere  $A_{D-2}=\frac{2 \pi^{\frac{D-1}{2}}}{\Gamma \left(\frac{D-1}{2}\right)}$, where $M$ represents the mass of the non-singular Black hole. The metric function $g^{tt}=f(r)$, whose sign determines gravitational trapping \cite{sa}, we plot to draw a sketch  to study the existence of black hole solutions.
The metric potential $f(r)$ is plotted with $r$ in Fig. (1) for $D=4$ and Fig. (2) for $D=10$. It is evident that both the extreme black hole and non-extreme black holes can be obtained for a given set of values of $r_g$ and $r_0$. 
We note that  extreme black hole exists for 
$r_g=1.0$ and $r_0=0.57$ in $D=4$ and $r_g=2.0$ and $r_0=1.57$ in $D=10$. In the first case no black hole exist for $r_g <1.0$  and the later case for $r_0>1.57$.
The photon radii are tabulated in Table-I for $D=4$ and Table-II for $D=10$. It is found that for  $D=4$, it increases with decrease of $\rho_0\sim1/r_0^2$ for a given mass but for a given $\rho_0$, photon radius is found to increase with mass. In $D=10$ dimensions as $\rho_0$ is decreases the photon radius decreases  then increases and decreases once again.
 In Fig (3) dimensional variation of the photon radius for $M=1$ is plotted for non-singular black holes with dimensions. The photon radius is maximum at $D=4$ and then decreases sharply as the dimensions is increases and remains constant.

\begin{figure}[t]
\centering \includegraphics[width=6 cm,height= 4 cm]{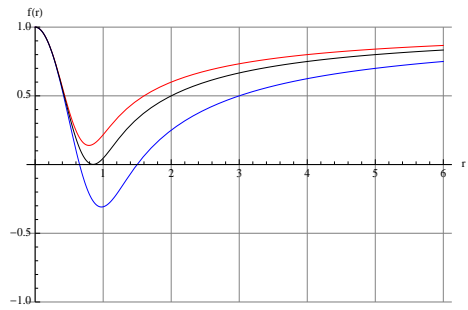}  
\caption{Radial variation of $f(r)$ for $r_g =$ 0.8 (Red), 1.0 (Black), 1.5 (Blue) for $r_0=0.57$ in $D=4$.}
\end{figure} 

\begin{figure}[]
\centering \includegraphics[width=6 cm,height= 4 cm]{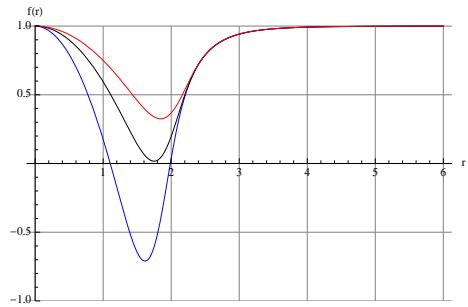}  
\caption{Radial variation of the metric function $f(r)$ in $D=10$ for $r_0= 1.1$ (Blue), $1.57$  (Black) and $2.0$ (Red) with $r_g=2$. }
\end{figure}

 \begin{figure}[b]
\centering \includegraphics[width=6 cm,height= 4 cm]{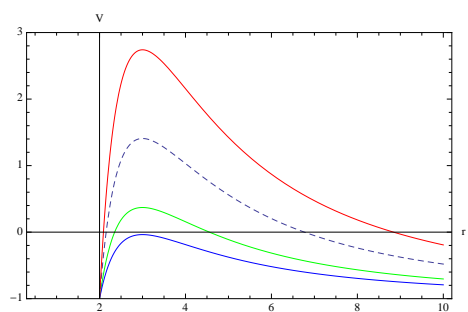}  
\caption{Radial variation of the potential for $J$= 5 (Blue), 6 (Green),  8 (Dashed), 10 (Red) in $D=4$ for non-singular BH.}
\end{figure}

The Lagrangian is given by
\begin{equation}
\label{26}
\mathcal{L} = \frac{1}{2} g_{AB} \dot{x}^{A} \dot{x}^B.
\end{equation}
where $\dot() = \frac{d}{d\tau}$ and $\tau$ is the affine parameter.
Expanding eq. (\ref{26}) we get
\begin{equation}
\label{27}
2\mathcal{L} =  -f(r) \dot{t}^2 + \frac{1}{f(r)}\dot{r}^{2} + ( r^2 \dot{\theta_1}^2 + sin^2 \theta_1 \dot{ \theta_2}^2 +  ... ...)
\end{equation}
To obtain trajectory of light path, we set  $\theta_i=\frac{\pi}{2}$ where $i=1, ... , D-3$ and $\theta_{D-2} $ is a free parameter. The momenta are given by
\[
P_t= \frac{\partial \mathcal{L} }{\partial \dot{t}}=  -f(r) \dot{t}, \; \; \; P_r=  \frac{\partial \mathcal{L} }{\partial \dot{r}}=  \frac{1}{f(r)} \dot{r},
\]
\begin{equation}
\label{28}
 P_{\theta_1} = \frac{\partial \mathcal{L} }{\partial \dot{\theta_1}}=  r^2\dot{\theta_1}, \; \; P_{\theta_2}= \frac{\partial \mathcal{L} }{\partial \dot{\theta_2}}=  r^2 sin^2 \theta_1 \dot{\theta_2 },... .
\end{equation}
Now as defined above, $\theta_i=\frac{\pi}{2}$, and at the equatorial plane $\theta_1=\frac{\pi}{2}$, 
\begin{equation}
\label{29}
\frac{\partial \mathcal{L} }{\partial \dot{t}}=constant 
\end{equation}
and we determine the energy ($E$) and angular momentum ($J$) at $r\rightarrow \infty$ as
\begin{equation}
\label{30}
f(r) \dot{t}=E,    \;\; P_{\theta_{D-2}}= r^2 \dot{\theta_{D-2}} = J. 
\end{equation}

The Hamilton Jacobi equation is the most general method to find the geodesic equation of motion around black hole or a compact object, we adopt the technique to obtain the photon orbits. In higher dimensions we get
\begin{equation}
\label{33}
\frac{\partial S}{\partial \tau} =\mathcal{H} = - \frac{1}{2} g^{AB} \frac{\partial S}{\partial x^A} 
\frac{\partial S}{\partial x^B} 
\end{equation}
where 
$g^{AB} $ is the inverse of the metric and $S$ is the Jacobian. The Jacobian is given by
\begin{equation}
S= \frac{1}{2} m^2 \tau - E + J \theta_{D-2} +S_r(r) + \sum_{i=1}^{D-3} S_{\theta_i}  (\theta_i)
\end{equation}
where $S_r(r)$ and $S_{\theta_i}  (\theta_i)$ are functions of $r$ and $\theta_i$ respectively and $m$ is the mass of the test particle, it is zero for photon. The Hamilton-Jacobi eq. (\ref{33})  can be written as
\begin{equation}
\label{34}
r^4 \left(1- \frac{R_s}{r^{D-3}}\right)^2  \left(\frac{\partial S}{\partial \tau} \right)  = E^2 r^4 -r^2 \left(1- \frac{R_s}{r^{D-3}}\right) (\mathcal{K}+J^2)
\end{equation}
\begin{equation}
\label{35}
\sum_{i=1}^{D-3} \frac{1}{\Pi_{n=1}^{i-1} sin^2 \theta_n} \left( \frac{\partial S_{\theta_i}}{\partial \theta_i} \right)^2  =\mathcal{K} - \Pi_{i=1}^{D-3} J^2 cot^2 \theta_i
\end{equation}
where $\mathcal{K}$ is the Carter constant \cite{cart}. Using the above eq. (\ref{28}) in eq. (\ref{33}) we get the following 
\[
\dot{t} = \frac{ E}{f(r)}, \;\;\;\; \dot{ \theta}_{D-2}= \frac{J}{r^2 \Pi_{i=1}^{D-3} sin^2  \theta_i} ;
\]
\begin{equation}
\label{36}
r^2\dot{r} = \pm \sqrt{\mathcal{R}}, \; \; \; r^2 \sum^{D-3}_{i=1} \Pi_{n=1}^{i-1} sin^2 \theta_n \dot{\theta}_i = \pm\sqrt{\Theta_i } 
\end{equation}
in the above "+" and "-" sign corresponds to motion of photon in outgoing and incoming radial direction  and over dot represents derivative w.r.t to the affine parameter$\tau$. For the null curves the eqs.(\ref{35}) can be expressed as 
\begin{equation}
\label{37}
 \mathcal{R}(r)=  E^2r^4 -r^2 f(r) ( \mathcal{K}^2 + J^2), 
\end{equation}
\begin{equation}
\label{38}
 \Theta_i(\theta_i) =  \mathcal{K}  - \Pi_{i=1}^{D-3} J^2 cot^2 \theta_i.
\end{equation}
The characteristics of photon near the black hole can be defined by two impact parameters, which are functions of the constants $E$, $J$ and  $ \mathcal{K}$. For general orbit we define the impact parameters $\xi = \frac{J}{E} $ and $\eta= \frac{\mathcal{K}}{E^2}$. The boundary  of the shadow of a black hole can be estimated from the effective potential. The radial null geodesic  from eqs. (\ref{34}) and (\ref{36}) is given by
 \begin{equation}
\label{39}
 \left(\frac{dr}{d\tau} \right)^2 + V_{eff}= 0, 
\end{equation}
where $ V_{eff}$ is the effective potential,  for radial motion we obtain
\[
  V_{eff}= \frac{f(r)}{r^2} ( \mathcal{K}+ J^2) -   E^2
\]
\begin{equation}
\label{40}  
\; \; \; =  \frac{1}{r^2}  \left[ 1-  \left( \frac{r_g}{r} \right)^{D-3} \left(1-    e^{- \left(  \frac{r}{r_{*}}\right)^{D-1}} \right) \right] ( \mathcal{K}+ J^2) -   E^2.
\end{equation}
The effective potential  is identical to the classical equation describing the motion of a massless particle in a 1-dimensional potential $V(r)$ provided its energy is $\frac{1}{2} E^2$ (of course the true energy should be $E$),  but we use this form to obtain an expression for potential in our study.
 We plot radial variation of $V(r)$ in Fig. (4) in a four dimensional universe for singular as well as non-singular black hole. As the angular velocity increases the photons heading towards the black hole  are unstable. In Fig. 2, it is found that there is no difference of the behaviour of the potential. In Fig. (4) we plot radial variation of $V(r)$ for different angular momentum,   for a non-singular BH,  it is evident that as the angular momentum increases the photon can approach near to the BH unbounded
  
%\begin{figure}[b]
%\centering \includegraphics[width=6 cm,height= 4 cm]{bh0.eps}  
%\caption{Radial variation of the potential for $J$= 5 (Blue), 6 (Red), 10 (Green) for a singular black hole.}
%\end{figure}
%\begin{figure}
%\centering \includegraphics[width=6 cm,height= 4cm]{bh1.eps}  
%\caption{Radial variation of the potential for $J$= 10 for singular (Red) and non-singular black holes ( Dashed).}
%\end{figure}

The photon orbits are circular and unstable for a maximum value of the effective potential. The unstable circular orbit determines the boundary of the apparent shape and can be maximized. The maximal value of the effective potential corresponds to the circular orbits and the unstable photons satisfies
\begin{equation}
\label{41}
 V_{eff}{\Big |}_{r=r_p}=  \frac{dV_{eff}}{dr}{\Big |}_{r=r_p} = 0, \; \; \; \mathcal{R}(r)=  \frac{d \mathcal{R}(r)}{dr}{\Big |}_{r=r_c}=0
\end{equation}
The impact parameters are now related as 
Using eqs. (\ref{40})  and  (\ref{41}), we get
\[
\frac{f(r_p)}{r_p^2} (\mathcal{K} + J^2) - E^2 =0
\]
\begin{equation}
\label{42}
 \frac{r_p f'(r_p) -2 f(r_p)}{r_p^3}  (\mathcal{K} + J^2)  =0.
\end{equation}
In four dimensions the potential $V(r)$ is plotted in Fig. (4) with different  angular momentum ($J$) for $r_g=2$ and  $E=1$. The  particles are bounded for a radius $r<r_{min}$  and unbounded for the range $r_{min} <r<r_{max}$. The range of values can be determined from the sketch. It is found that $r_{min}$ decreases as angular momentum ($J$) increases. We note that the potentials for Schwarzschild black hole (singular) and that for non-singular black holes overlaps for a set of similar values of   $D$, $J$ and $E$.

\begin{table}
 \centering
  \begin{tabular}{|c|c|c|c|} \hline
    & $r_p$ in & $r_p$ in & $r_p$ in \\ 
$r_0$ & $M=2$  &  $M=5$  & $M=10$ \\  \hline 
0.5 & 0.8757   & 1.0192 & 1.1532          \\ \hline
0.6 & 1.0221   & 1.1851& 1.3389             \\ \hline
0.8 & 1.3079   & 1.5053 & 1.6958                \\ \hline
1.0 & 1.5880  & 1.8142 & 2.0383                \\ \hline
1.2 & 6.0000  & 2.1150 & 2.3702               \\ \hline
1.5 & 6.0000 & - & -                \\ \hline
1.8& 5.990   & -& -            \\ \hline
2 &2.9921 & - & -             \\ \hline
\end{tabular}
\caption{The variation of the photon radius ($r_p$) in $D=4$ with $r_0=\sqrt{\frac{(D-1)(D-2)}{4\rho_0}}$ and the mass of the BH. }
 \label{tab:1}
\end{table}

\begin{table}
 \centering
  \begin{tabular}{|c|c|c|} \hline
  & $r_p$ in & $r_p$ in \\ 
$r_0$ & $M=5$   & $M=10$ \\  \hline 
0.5 & 0.9369   & 1.0746          \\ \hline
0.6 & 1.0035  & 1.0002            \\ \hline
0.7 & 1.0641 & 0.9863           \\ \hline
0.8 & 1.1203   & 1.0613               \\ \hline
0.84 &  1.1417 & 24.6008            \\ \hline
0.85 & 1.1470       & 0.9472          \\ \hline
0.89 & 1.178  & 1.5069            \\ \hline
0.9 &  1.1730      & 25.2259      \\ \hline
0.91 & 1.1781        & 0.9598      \\ \hline
0.95 &     1.1983   & 1.1687     \\ \hline
1.0 & 1.2229  & 0.9772               \\ \hline
1.2 & 1.8426  & 1.0105             \\ \hline
2 & 3.5942 & 6.9106               \\ \hline
\end{tabular}
\caption{The variation of the photon radius ($r_p$) in $D=10$ with $r_0=\sqrt{\frac{(D-1)(D-2)}{4\rho_0}}$ and the mass of the BH. }
 \label{tab:1}
\end{table}

\begin{figure}[t]
\centering \includegraphics[width=6 cm,height= 4 cm]{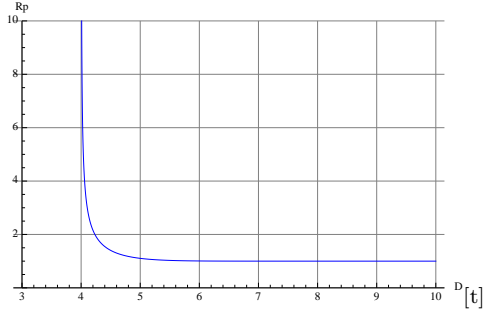}[t]
\caption{Dimensional variation of the photon radius  for $M=1$ for a non-singular BH}
\end{figure}
We draw the shadow contour of non-singular black hole in Fig. (8) for a given value of $\rho_0$ (say, $0.16$ unit) in all dimensions. It is shown that as the spacetime dimensions increases the radius of the shadow decreases.
\begin{figure}
\centering \includegraphics[width=5 cm,height= 5 cm]{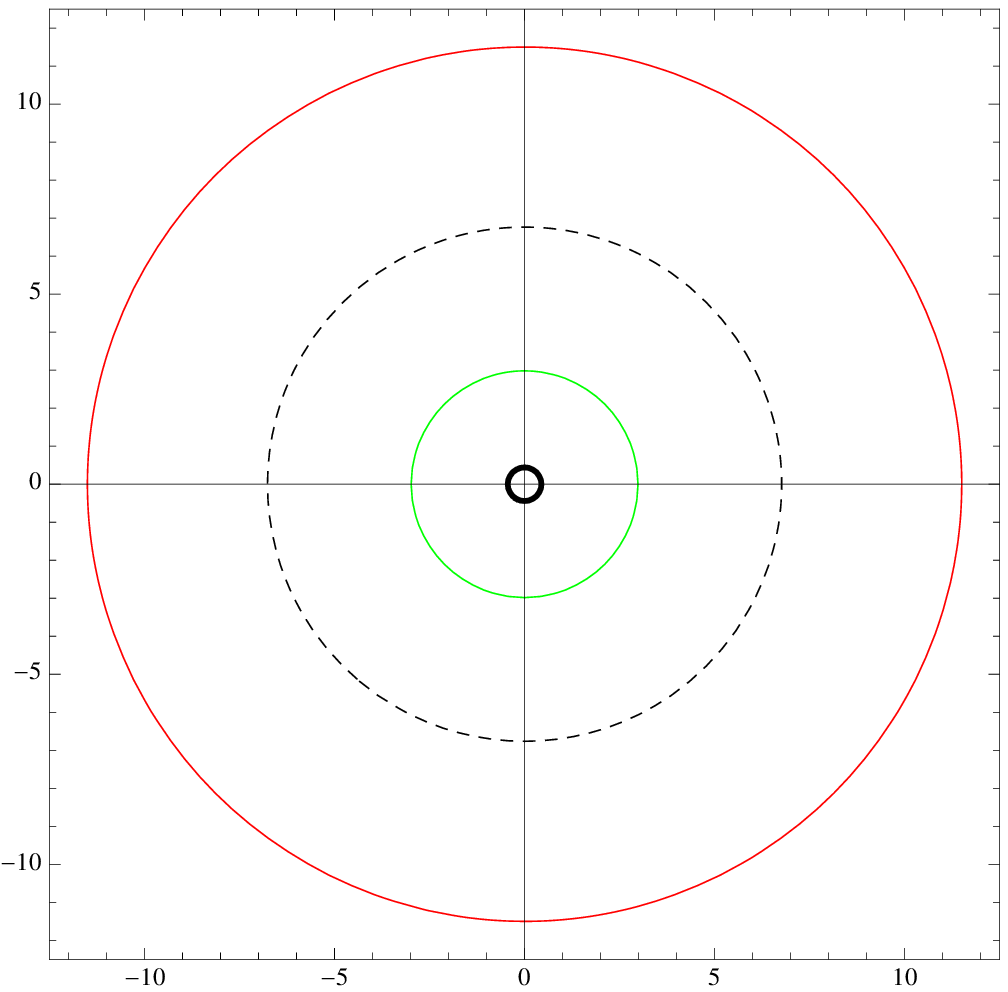} 
\caption{Contour plot for  an object having  $M=2 M_{\odot}$ with $\rho_0=0.16$ unit for $D=4$ (Red),  $D=5$ (Dashed) and $D=6$ (Green), $D=8$ (Thick)}
\end{figure}

\begin{figure}[t]
\centering \includegraphics[width=5 cm,height= 5 cm]{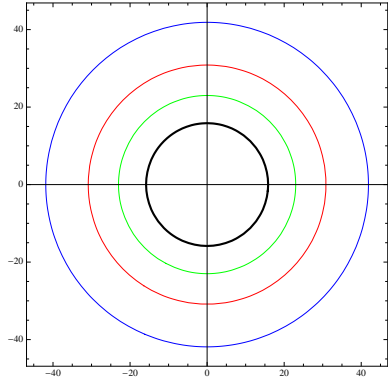} 
\caption{Contour plot for  an object for  $\rho_0=0.04$ unit in $D=4$  with   $M=2 M_{\odot}$ (Black), $M=4 M_{\odot}$ (Green), $M=6 M_{\odot}$ (Red),  $M=10 M_{\odot}$ (Blue) }
\end{figure}

\section{Effective potential and shadow behaviour}
The effective potential of the Schwarzschild-Tangherlini black holes exhibits a maximum for the photon sphere radius $r_p$ corresponding to the real and the positive solution of the constraint obtained from eq. (\ref{42}),
\begin{equation}
\label{43}
 r_p f'(r_p) - 2 f(r_p)  =0.
\end{equation}
Defining impact parameters $\eta$ and $\xi$ that are functions of the energy $E$, angular momentum $J$ and the Carter constant $\mathcal{K}$ as 
\begin{equation}
\label{43a}
\xi = \frac{J}{E}, \; \; \; \; \eta= \frac{\mathcal{K}}{E^2}
\end{equation}
we get from  eq. (\ref{41}) corresponding to $\frac{V_{eff}}{E^2}=0 $ and $\frac{\mathcal{R}}{E^2}=0 $, the following 
\begin{equation}
\label{44}
\eta+\xi^2=\frac{r_p^2}{f(r_p)}, \; \; \eta+\xi^2=\frac{4r_p^2}{r f'(r_p)+ 2f(r_p)}.
\end{equation}
Now we obtain 
\begin{equation}
\label{45}
\eta+\xi^2=\frac{5 r_p^2}{r_p f'(r_p)+ 3f(r_p)},
\end{equation}
where the right hand side corresponds to $\frac{r_p^2}{f(r_p)}$, the observer's frame the shadow can be described properly making use of the celestial coordinates $\alpha$ and $\beta$ as introduced earlier \cite{22}. Following the definition introduced by Subrahmanyan as follows
\begin{equation}
\label{46}
\alpha = \lim_{r_p \rightarrow \infty} \left( \frac{ r_p P^{\theta_{D-2}}}{P^t}\right), \;\; \;  
 \beta_i= \lim_{r_p \rightarrow \infty} \left( \frac{ r_p P^{\theta_{i} }}{P^t} \right),
\end{equation}
\[
where  \;\;\; i =1, ...(D-3).
 \]
For an observer on the equatorial plane, these equations reduced to
\begin{equation}
\label{45}
\eta+\xi^2= \alpha^2+\beta^2= \frac{r_p^2}{f(r_p)}
\end{equation}
the radius of the shadow is $R_{bhs}=\frac{r_p}{\sqrt{f(r_p)} }$. The form of $f(r)$ is complex and therefore we study numerically. The photon radius depends on the  dimensions. The photon radius  is plotted in Fig. 4, it is evident that  as the mass of the black hole increases the radius decreases. It is maximum in $D=4$ but decreases sharply as the dimension increases but almost constant with the increase in dimension. 
The fig (9) shows that as the mass increases the radius of the shadow also increases.

\section{Discussion}

We obtain non-singular black hole (NSBH) solutions in the higher dimensional Einstein's general theory of gravity (GR) and found that the methods  in GR can be adopted also in Rastall gravity. Considering a specific exponential form of the energy density we obtain NSBH which reduces to the Dymnikova NSBH solution \cite{18} obtained in the usual four dimensions  ($D=4$). In $2+1$ dimensions no black hole solution exists. However,   a non-rotating NSBH solutions obtained by Dymnikova in four dimensional GR can be accommodated in higher dimensions. 
We obtained  NSBH in a vacuum described by $T^t_t+T^r_r=0$ with $p_{\perp} <0$  (where $i=1, ..., D-2$) near the center indicating requirement of exotic matter which however extends up to certain height thereafter  $p_{\perp} >0$ for $r > \left( \frac{D-2}{D-1} \right)^{\frac{1}{D-1} } r_{*}$ in GR.  But in the Rastall gravity $p_{\perp} >0$  for  $r > \left( \frac{2-D+ 2 \lambda D}{(D-1)(2\lambda-1)} \right)^{\frac{1}{D-1} } r_{*}$ when $\lambda \neq  \frac{D-2}{2D}$  
and thereafter at a large distance it vanishes because  the pressure decreases rapidly $ i.e., $ exponentially.  Both in GR and modified gravity it  indicates existence of exotic matter near the center of the NSBH but in the later case the Rastall parameter plays an important role in determining the distance from the centre where the normal matter exists in the tangential direction.
In the usual four dimensions at the center of the NSBH  in the modified theory  we get the following estimations $\rho (r=0)= 2(1-4\lambda) \rho_0$ and tangential pressure $P_{\perp}= -2(1-4\lambda)$ which are determined by the Rastall parameter $\lambda$. It is evident that exotic matter at the center of the black hole requires both in GR (as $\lambda=0$) as well as in Rastall theory with the lower limiting value of the Rastall parameter $\lambda <  \frac{1}{4}$.
 The tangential pressure is negative it indicates NSBH  surrounded by exotic matter in Rastall gravity, existence of BH with exotic matter also reported in \cite{R1}.   
 Thus NSBH is realized with both the central radial pressure and tangential pressure negative and equal initially but an anisotropy in pressure develops away from the center in Rastall gravity with normal matter thereafter when  
 $r > \left(\frac{D-2+2\lambda D}{(1-2 \lambda)(D-2)} \right)^{\frac{1}{D-1}} r_{*}$. 
   
However, we  note a new and interesting result in Rastall theory that permits a NSBH  with normal matter in the usual four and in higher dimensions when $\lambda =\frac{D-2}{2D}$, which however is not permitted in GR.
In  It is also noted that   away from the center at a large distance, the tangential pressure remains positive definite at a maximum radial distance which is $P_{\perp}=  (1-2 \lambda) \frac{D-1}{D-2} \frac{r^{D-1}}{r_{*}^{D-1}} 
 \rho_0 e^{- \frac{r^{D-1}}{r_{*}^{D-1}}}$. Thus one gets a physically realistic NSBH  for $\frac{1}{4}<\lambda < \frac{1}{2}$ in $D=4$ dimensions.
 For $r\rightarrow \infty $, both the energy density and pressure vanishes asymptotically. Thus the Rastall gravity has rich structure which unearth the structure of non-singular black hole even with normal matter for a restricted domain of the Rastall parameter depending on the embedding spacetime dimensions.
 Thus, we see that for $\lambda \neq 0$ the Rastall theory  plays an important role leading to distinct solutions relative to GR. 
 
 The sketch of the potentials permissible in the theory are plotted in Figs. (1) and (2), which show that both extreme and non-extreme black holes exist.

The contour plots in Fig. (5) for NSBH shows that the circular shadow radius decreases as the spacetime dimension is increased for a given mass. The  circular shadow radius in Fig. (6)  show that the radii increases with the mass of the compact objects for a given dimensions. The rotating NSBH will be taken up elsewhere.

{\bf Acknowledgment}
The author would like to thank IUCAA , Pune and IUCAA Centre for Astronomy Research and Development (ICARD), NBU for extending research facilities and  North Bengal University for a research grant. BCP acknowledge the suggestions and constructive criticism of the anonymous Referee.  \\

{}
\end{document}